# Interfacial Thermal Transport in Si/SiC and SiC/diamond Heterostructures: Effects of Amorphous Interlayers and SiC Polytypes

Pedram Mirchi[a,1], Shirin Sabokdast[b], Ali Rajabpour[c,2]

[a] Department of Mechanical Engineering, K. N. Toosi University of Technology, Tehran, Iran

[b] School of Quantum Physics and Matter, Institute for Research in Fundamental Sciences, Tehran, Iran (IPM)

[c] Mechanical Engineering Department, Imam Khomeini International University, Qazvin, Iran

**Corresponding Authors:**

[1] **Pedram Mirchi** — E-mail: pedrammirchi@gmail.com

[2] **Ali Rajabpour** — E-mail: rajabpour@eng.ikiu.ac.ir

## Abstract

This study examines phonon-mediated heat transfer across Si/SiC and SiC/diamond interfaces using non-equilibrium molecular dynamics simulations, emphasizing the influence of SiC polytypes and amorphous interlayers. For sharp interfaces, 4H-SiC exhibits considerably higher interfacial thermal conductance (ITC) than 3C-SiC, due to its broader active phonon spectrum and superior spectral matching with Si. While amorphous layers generally reduce ITC, a key observation is that an ultrathin 0.5-nm amorphous SiC (aSiC) layer can enhance heat transport in the Si/3C-SiC system: the ITC increases from 613 MW/m²·K (sharp) to 716 MW/m²·K, demonstrating a phonon-bridge effect. VDOS (vibrational density of states) analysis confirms that optimized ultrathin aSiC layers improve vibrational overlap and open additional phonon-transport channels. In contrast, thicker or silicon-rich amorphous layers significantly suppress ITC through enhanced inelastic phonon scattering. For SiC/diamond interfaces, any amorphous layer — particularly aSi — causes severe ITC degradation, highlighting the need for sharp, defect-free bonding to exploit diamond's high thermal conductivity.

# 1. Introduction

Recent developments in power electronics and materials science have led to a substantial rise in the power density of semiconductor devices. This increase is accompanied by excessive heat flux, which significantly raises the junction temperature in high-power electronics, reducing electron transport efficiency and posing a serious risk to device reliability. Thanks to their high breakdown voltages, excellent electron mobility, and fast switching capabilities, III-V semiconductors are widely employed in the fabrication of high-power, high-frequency devices for applications such as radio-frequency communications, medical equipment, and industrial manufacturing. Nonetheless, thermal management remains a major challenge for III-V devices. For instance, recent reports indicate that local power densities in III-V-based high electron mobility transistors (HEMTs) can reach roughly 100 kW/cm$^2$, and the high channel temperatures (~200 ℃) may severely limit device performance [1-3]. Therefore, effective cooling of high-power devices remains a critical challenge that must be addressed. To enhance thermal management in electronic devices, recent studies have explored reducing the overall thermal resistance through the use of high-thermal-conductivity substrate materials, including diamond (C), BAs, and silicon carbide (SiC) [4-8].

GaN-on-diamond structures have been proposed, taking advantage of diamond's exceptionally high thermal conductivity (>1000 W/m·K) for efficient heat removal from device junctions [9, 10]. However, direct bonding or heteroepitaxial growth of GaN on diamond is hindered by lattice and thermal expansion mismatches [11].

While diamond has the highest isotropic thermal conductivity among bulk materials, its high cost, small wafer size, and integration challenges limit its use [12-14]. Graphite and carbon-based nanomaterials like graphene and carbon nanotubes show strong anisotropy or reduced κ when combined, restricting their practical application [15, 16].

Silicon carbide (SiC) is a key material in a variety of emerging technologies, including power electronics, optoelectronics, and quantum computing [17-19]. Power devices built on SiC have the potential to revolutionize power electronics by replacing silicon-based technologies, thanks to their high switching speeds, low energy losses, and elevated blocking voltages [20]. High thermal conductivity (κ) materials like hexagonal SiC (6H, 4H) and AlN have room-temperature c-axis κ values of roughly 320–350 W/m·K, which are lower than metals such as silver and copper [21, 22]. The previously cited κ of 490 W/m·K for 6H-SiC originated from Slack's 1964 measurements [23, 24], but more recent TDTR experiments corrected it to ~320 W/m·K [5, 21, 25], matching DFT predictions for perfect single-crystal 6H-SiC [26]. This agreement confirms the high quality of commercially available 6H-SiC. Compared to the widely studied hexagonal SiC (6H and 4H), cubic SiC (3C) is less understood, yet it offers superior electronic properties and higher thermal conductivity [17, 20]. MOSFETs based on 3C-SiC achieve the highest channel mobility among all SiC polytypes, significantly reducing power consumption, and 3C-SiC can be directly grown on silicon [20].

Zhe Cheng et al. [27] reported that wafer-scale 3C-SiC exhibits isotropic thermal conductivity exceeding 500 W/m·K at room temperature, ranking second only to diamond among large crystals. They also found that corresponding 3C-SiC thin films show record-high in-plane and cross-plane thermal conductivities, surpassing diamond films of similar thickness. This exceptional thermal performance was attributed to the high crystal quality and purity, which minimizes defect-phonon scattering. Furthermore, 3C-SiC can be epitaxially grown on silicon, with 3C-SiC/Si interfaces exhibiting among the highest measured thermal boundary conductances, highlighting its potential as a wide-bandgap semiconductor for next-generation power electronics, both as active layers and thermal management substrates. Xinlong Zhao et al. used molecular dynamics simulations to systematically examine thermal transport at Si/diamond interfaces in 3D ICs, highlighting the critical influence of atomic composition and interfacial structure on ITC. Comparing Si/diamond, SiC/diamond, and diamond/diamond interfaces revealed that material mismatches dominate ITC reduction, supported by phonon density analyses. Introducing a 1 nm SiC transition layer at the Si/diamond interface increased ITC to 701.1 MW/m²·K —~50% higher than pure Si/diamond and over twice the DMM prediction. A 3D IC thermal model further demonstrated that interface ITC strongly affects device peak temperatures, providing guidance for interface optimization in high-performance 3D ICs [28].

Recent studies have shown that interfacial thermal transport can be effectively tuned through various interface engineering strategies in different material systems. For example, molecular dynamics simulations have demonstrated that the ordering of polypropylene chains, an increase in matrix density, and graphene functionalization can enhance the interfacial thermal conductivity at the graphene/polypropylene (PP) interface by strengthening phonon coupling [29].

In another study on the graphene/silicon system, nitrogen doping was found to significantly reduce the interfacial thermal resistance, and when arranged in an ordered distribution, it improved heat transfer by creating additional phonon transport channels [30]. Similarly, atomic-scale surface engineering, such as introducing roughness or amorphous layers in the Si substrate, has been shown to effectively regulate interfacial heat transfer in $MoS_2$/Si heterostructures by modifying phonon coupling and scattering [31]. Moreover, recent work combining machine learning techniques with NEMD simulations demonstrated that optimizing defect distributions can significantly enhance the interfacial thermal conductivity in graphene/h-BN heterostructures by generating optimal atomic configurations [32].

Previous research has examined ITC of SiC/Si interfaces. Xu et al. showed that introducing nanostructures at the 4H-SiC/Si interface can increase ITC from 300 to 1000 $MW/m^2\cdot K$ [33], while Guo et al. reported 293 $MW/m^2\cdot K$ for 4-in. 4H-SiC/Si wafers fabricated via surface-activated bonding after annealing at 750 °C [34]. Although nanostructuring and epitaxial growth can enhance ITC, they may also introduce interface disorder that degrades device performance. In contrast, introducing interlayers such as amorphous $SiO_2$ or Si, which can act as phonon bridges, has been proposed as a practical approach for tuning ITC [35-38]. However, phonon scattering in these amorphous layers can reduce interfacial thermal transport [39], highlighting the importance of understanding their thermal conductivity mechanisms for optimizing SiC/Si thermal management. This molecular dynamics study [40] shows that inserting an ultrathin amorphous Si (aSi) interlayer can significantly enhance ITC at the Si/diamond interface, with a 0.5 nm layer increasing ITC by over 38% by reducing phonon mismatch. However, further increasing the aSi thickness lowers ITC, while higher temperatures enhance ITC due to stronger inelastic phonon scattering, indicating that precise control of interlayer thickness is key for thermal management in Si/diamond devices. Yang Zhang et al., using molecular dynamics simulations, demonstrated that SiC/Si wafers with thin amorphous $SiO_2$ or Si interlayers can significantly influence ITC. They found that a 0.3 nm $aSiO_2$ layer can increase ITC up to 994.7 $MW/m^2 \cdot K$, which is 54.4% higher than that of SiC/Si interfaces without an amorphous layer, whereas aSi layers reduce ITC. This enhancement is attributed to better overlap of vibrational density of states and the creation of new phonon transport channels, and the temperature-dependent ITC results highlight the crucial role of interlayer type and thickness in optimizing thermal management in SiC/Si wafers [41].

In this study, non-equilibrium molecular dynamics (NEMD) simulations are employed to systematically investigate ITC of Si/SiC and SiC/diamond interfaces. The influence of different SiC polytypes (3C and 4H) as well as the presence of ultrathin and thick amorphous interlayers (aSi, aSiC, and mixed aSi-aSiC) on interfacial heat transport is examined. Furthermore, spectral thermal conductance, Phonon Participation Ratio (PPR) and VDOS analyses are performed to clarify the phonon transport mechanisms and the role of amorphous layers as potential phonon bridges. Furthermore, quantum-correction effects were applied to the ITC calculations and the results were compared with those obtained from classical simulations. These results provide theoretical insights for interface engineering and thermal management optimization in SiC-based high-power electronic devices.

## 2. Models and simulation method

The atomic configurations of the investigated heterostructures were constructed to include both sharp interfaces and interfaces containing intermediate amorphous layers. For the Si/SiC interfaces, the intermediate layers included aSi (0.4 nm), aSiC with thicknesses of 0.5, 1.2, and 1.8 nm, and a composite aSi/aSiC interlayer (0.8 nm). For the SiC/diamond interfaces, the same amorphous configurations were considered, except that the composite interlayer consisted of aC/aSiC instead of aSi/aSiC. These configurations allow a direct comparison between ideal interfaces and interface-engineered structures incorporating transition layers. The different atomic configurations used in this study are illustrated in Figure 1.

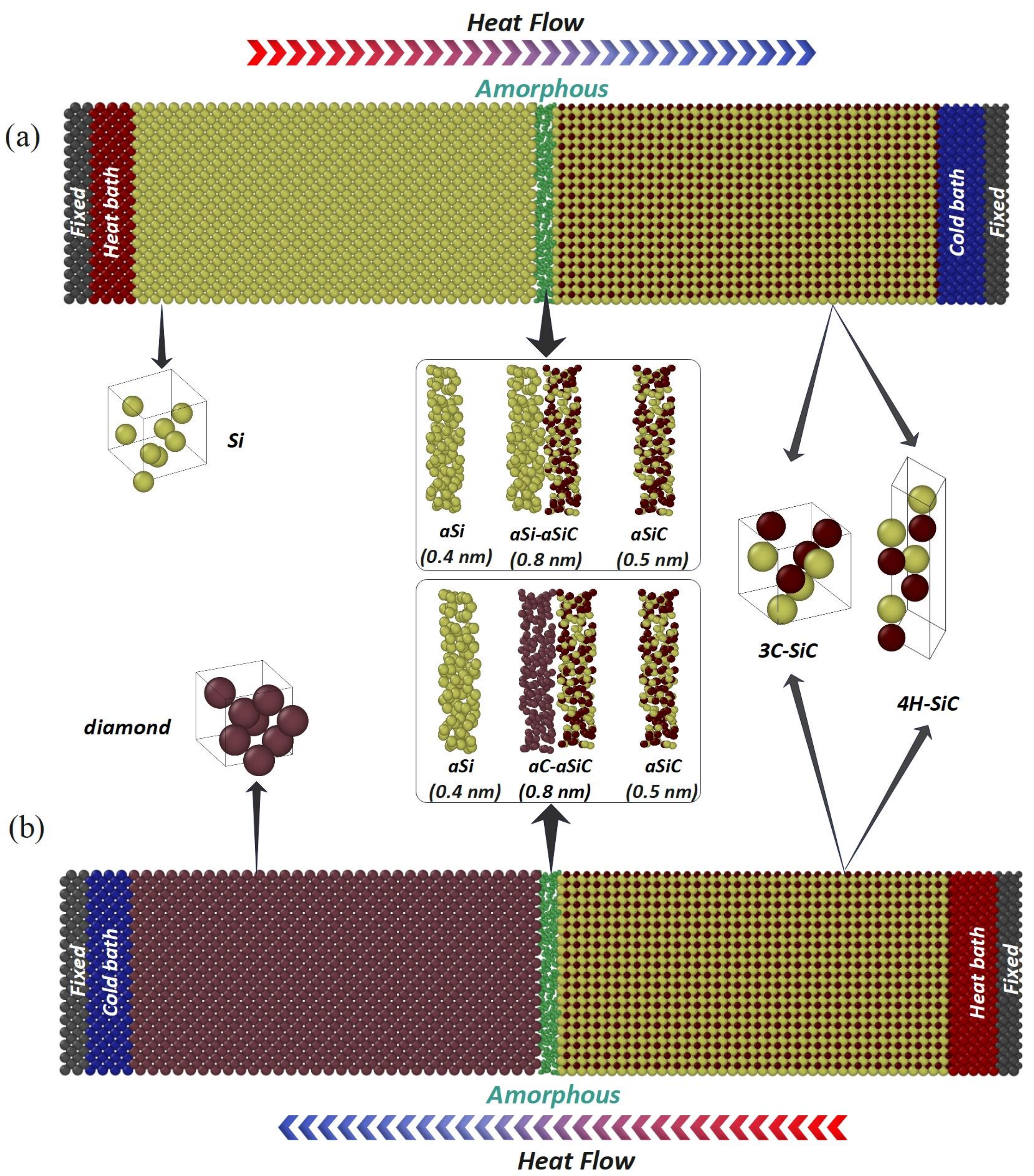

Figure 1: (a) Atomic structure of the Si/aSi/SiC, Si/aSiC/SiC, and Si/aSi-aSiC/SiC interfaces and (b) Atomic structure of the SiC/aSi/diamond, SiC/aSiC/diamond, and SiC/aC-aSiC/diamond interfaces and the NEMD simulation setup for calculating ITC.

Si, diamond, 3C-SiC, and 4H-SiC were arranged so that the simulation box x-axis coincided with the crystallographic [001] direction, placing the (001) planes parallel to the YZ plane. The two SiC crystal phases, 4H-SiC and 3C-SiC, were independently constructed and analyzed at the interface. All structural parameters and crystallographic orientations used in the simulations for 4H-SiC, 3C-SiC, diamond, and Si are summarized as follows.

Si and diamond have a cubic crystal structure with space group Fd-3m. The lattice parameter of Si is a = b = c = 5.444 Å, whereas diamond has a = b = c = 3.55 Å. The cubic polytype 3C-SiC belongs to the F4-3m space group with lattice parameters a = b = c = 4.354 Å. In contrast, 4H-SiC exhibits a hexagonal crystal structure with space group $P6_3mc$, characterized by lattice parameters a = b = 3.076 Å and c = 10.071 Å.

To minimize finite-size effects and ensure appropriate in-plane lattice compatibility, commensurate supercells were constructed using integer multiples of the SiC, diamond, and Si unit cells. Subsequently, an energy minimization procedure was carried out to fully relax both the simulation box dimensions and the atomic coordinates. This step reduces the initial stresses and excess potential energy in the system, allowing the atomic structure to settle into a mechanically stable configuration.

During the construction of the heterostructure, the lateral dimensions of both 3C-SiC and Si were set to 4.3 nm × 4.3 nm along the y and z directions, and periodic boundary conditions were applied in these directions. The lengths along the x-axis were set to 10.4 nm for both Si and 3C-SiC, resulting in a total simulation length of 20.8 nm along the x-axis. The heterostructure was modeled with lateral dimensions of 4.5 × 4.2 nm under periodic boundary conditions, while Si and 4H-SiC each had a length of 10.2 nm along the x-axis, giving a total simulation length of 20.4 nm. Also, the simulation cell dimensions were 5.3 nm × 5.3 nm in the y and z directions, while the x-direction length was set to 10 nm for both 3C-SiC and diamond, giving a total length of 20 nm. For the 4H-SiC/diamond heterostructure, the simulation cell had lateral dimensions of 4.2 × 4.6 nm, while both materials were 10 nm long along the x-axis, resulting in a total length of 20 nm. This overall size was considered sufficiently large to eliminate finite-size effects [41].

The model construction and NEMD simulations were performed using the Large-scale Atomic/Molecular Massively Parallel Simulator (LAMMPS) package [42]. The interatomic interactions were modeled using the Tersoff potential, which has been widely validated for Si, SiC, diamond, Si/SiC, and SiC/diamond systems in previous thermal transport studies [23, 28, 43]. A single consistent set of potential parameters was employed throughout all simulations to ensure a uniform description of interatomic interactions across all interface configurations. The accuracy of this potential in reproducing phonon dispersion relations and its consistency with experimental heat transport data are further detailed in Section S1 of the Supporting Information.

Once the simulation box was set, the next task was to position the atoms. Before further operations, a brief energy minimization step was taken to remove atomic overlaps and physically implausible configurations. Subsequently, structural relaxation was performed in the isothermal-isobaric (NPT) ensemble at 300 K for 250 ps to achieve thermodynamic equilibrium regarding temperature and pressure. Throughout all simulations, the time step was set to 0.25 fs. After the NPT equilibration, the system was brought to further equilibrium in the canonical (NVT) ensemble at 300 K for an additional 250 ps to achieve maximum uniformity in the distribution of velocities and for the system to become thermally stable. To carry out the heat transport simulation, the simulation cell was partitioned along the X direction into four different zones: fixed, hot, cold, and active. The fixed regions, which had a 5 Å thickness at each end of the system, had frozen atoms. The hot and cold zones were positioned respectively at 320 K and 280 K and were maintained by a Langevin thermostat, leading to a temperature difference of 40 K across the system. The heat transfer region was segmented into 20 distinct sections for precise temperature profile analysis. The NEMD simulation ran for 1000 ps to attain a non-equilibrium steady state, followed by an additional 1000 ps for data collection.

The ITC was computed from the following equation [44, 45]:

$$ITC = \frac{dE}{dt.A.\Delta T} \tag{1}$$

Here $\Delta T$ is the temperature gradient present across the two sides of an interface. $\frac{dE}{dt}$ is the rate of energy change with time, and $A$ is the cross-sectional area of the interface.

To ensure statistical reliability, each simulation was independently repeated four times with different initial velocity distributions, and the reported ITC values represent the average of these runs. The associated uncertainty, indicated by error bars in all figures, remained within ±3% across all configurations.

To explain the impact of the amorphous interlayer on heat transfer across the SiC/Si heterostructure interface, the VDOS of the interfacial region was analyzed for two cases: a sharp Si/SiC interface and a Si/aSiC (0.5 nm)/SiC structure. Considering that atoms located in the vicinity of the interface play the most significant role in ITC, the VDOS was obtained from a region up to 1 nm from the interface.

The VDOS was acquired through the Fourier transform of the velocity autocorrelation function (VACF) [46]:

$$VDOS(\omega) = \frac{1}{\sqrt{2\pi}} \int_{-\infty}^{\infty} e^{-i\omega t} \frac{\langle v(t).v(0)\rangle}{\langle v(0).v(0)\rangle} dt \tag{2}$$

In this expression, $\mathrm{VDOS}(\omega)$ represents the vibrational density of states at frequency $\omega$; $v(t)$ and $v(0)$ denote the atomic velocities at times $t$ and 0, respectively, and the angle brackets indicate ensemble averaging.

With the detailed spectral information, materials and interfaces can be designed to maximally enhance thermal transport properties. Spectral decomposition of thermal conductance calculations further provides insights into ITC by revealing the dominant scattering mechanisms associated with certain phonon modes, and therefore, deepens the understanding of ITC [11]. To further refine the estimation of phonon-mode contributions to heat transport across the Si/SiC interface, the spectral thermal conductance was evaluated using a spectral heat flux decomposition method [47, 48] based on NEMD simulations. The spectral heat current $q(\omega)$, transferred from the left side of the interface to the right side, can be expressed as follows:

$$q_{L\to R}(\omega) = \frac{2}{A} Re \sum_{i\in L} \sum_{j\in R} \int_{-\infty}^{\infty} \langle F_{ij}(t).\, v_i(0)\rangle\, e^{i\omega t} dt \tag{3}$$

Here, A indicates the cross-sectional area normal to the direction of heat flow; t is the time correlation between the forces and the velocities; $F_{ij}$ denotes the interatomic force between the atoms to the left and right of the interface; and $v_i$ is the velocity vector of the atom $i$ on the left side of the interface. The spectral thermal conductance $g(\omega)$ can be computed using these definitions as:

$$g(\omega) = \frac{|q_{L\to R}(\omega)|}{\Delta T} \tag{4}$$

where $\Delta T$ denotes the temperature discontinuity at the interface.

In classical molecular dynamics (MD), phonons follow Maxwell–Boltzmann statistics, which can overpopulate high-frequency modes when the simulation temperature is lower than the Debye temperature. Since the Debye temperatures of Si, SiC, and diamond (≈645 K, 1200–1300 K, and

>2000 K) are significantly higher than the simulation temperature (300 K), classical MD may overestimate the ITC.

Therefore, a quantum correction factor was applied to the spectral thermal conductance to reduce the contribution of high-frequency phonons and obtain a more realistic ITC [40, 41, 49, 50].

$$a = \frac{x^2 e^x}{(e^x - 1)^2} \quad (5)$$

In this relation, $x = \hbar\omega/k_B T$, where $\hbar$ denotes the reduced Planck constant, $k_B$ is the Boltzmann constant, and $T$ represents the system temperature.

The amorphous structures were generated by randomly placing Si and SiC atoms within the simulation cell. After an initial energy minimization step to eliminate atomic overlaps, the system was heated to 6000 K and equilibrated for 80 ps under the NVT ensemble to ensure complete melting. Subsequently, the system was gradually cooled from 6000 K to 300 K, allowing the formation of a relaxed amorphous structure at room temperature. The densities of the generated amorphous structures after equilibration were approximately 2.29 g/cm$^3$ for aSi and 2.88 g/cm$^3$ for aSiC, which are consistent with previously reported values for amorphous configurations [41, 51]. Details regarding the structural analysis and radial distribution functions (RDF) plots are provided in the Supporting Information S2.

## 3. Results and discussion

### 3.1. Spectral and accumulated ITC analysis of Si/SiC interfaces

To investigate the ITC at the Si/SiC in the presence of an amorphous layer, we first examined heat conduction without the amorphous layer. For constructing the Si/SiC heterostructure model, two SiC polytypes, 4H-SiC and 3C-SiC, were selected. These models allow for identifying the optimal crystal structure that provides an effective value for the ITC. The energy profiles extracted from the Langevin thermostat for the heterogeneous Si/3C-SiC and Si/4H-SiC structures are shown in Figures 2(a) and 2(b), respectively. The heat flux was obtained by averaging the rates of energy input and output at the heat source and heat sink regions. Figures 2(c) and 2(d) display the steady-state temperature profiles. As can be observed, a temperature drop occurs near the interface. The interfacial temperature discontinuity ($\Delta$T) was calculated by linearly extrapolating the temperature gradients on both sides of the interface up to the boundary, following the approach proposed by Chen et al. [52]

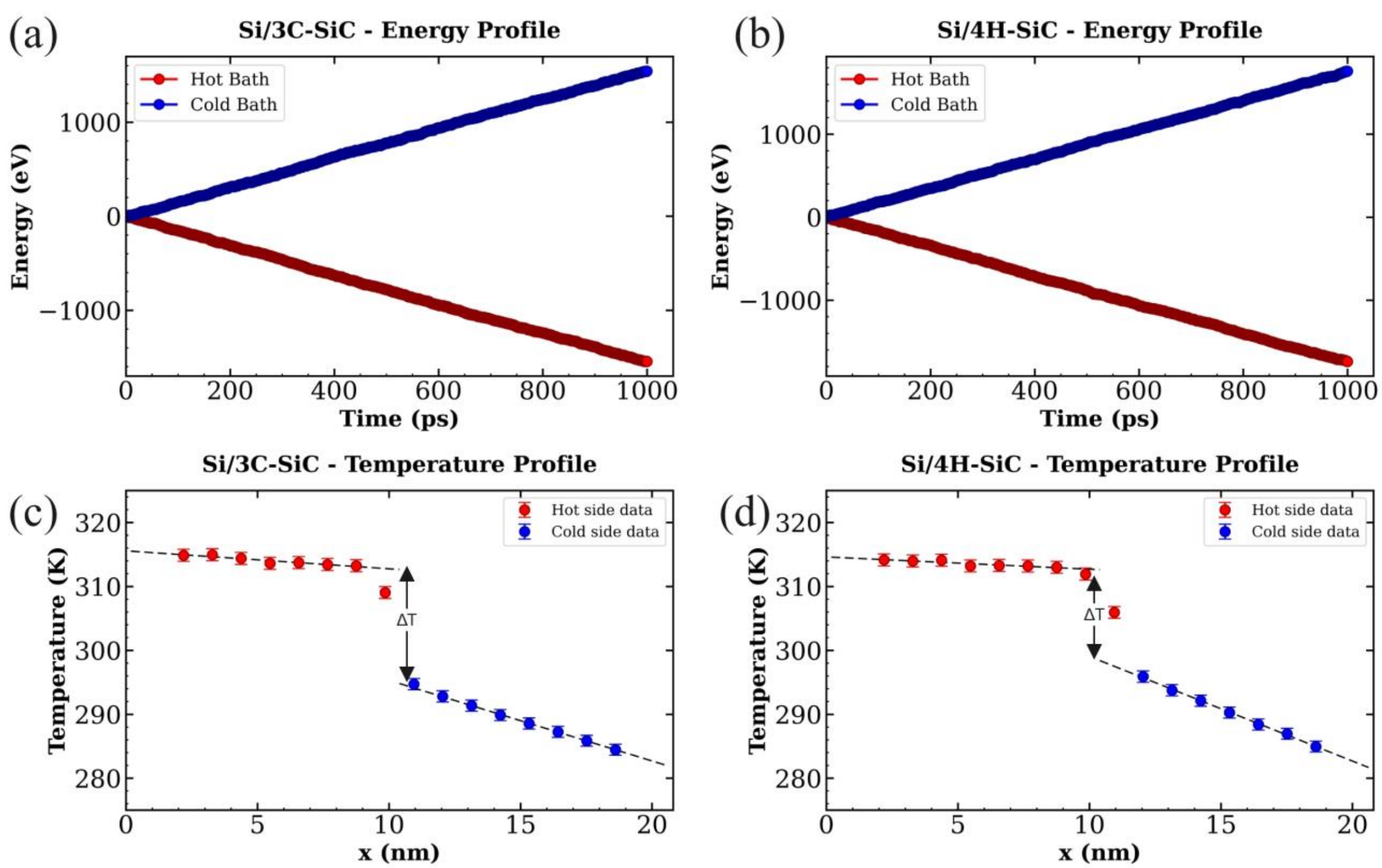

Figure 2. (a) Cumulative energy profile of thermostats during NEMD simulations for the Si/3C-SiC heterostructure; (b) same for Si/4H-SiC; (c) steady-state temperature profile for Si/3C-SiC; (d) steady-state temperature profile for Si/4H-SiC.

The spectral interfacial thermal conductance analysis of the Si/SiC interface indicates that the frequency distribution of phonons contributing to heat transfer strongly depends on the crystal structure (polytype) of SiC. The spectral thermal conductance results for the Si/3C-SiC and Si/4H-SiC interfaces are presented in Figure 3(a–d).

As shown in Figure 3(a), the Si/3C-SiC interface exhibits a dominant and broad peak within the 5–10 THz frequency range, with a maximum spectral thermal conductance of approximately 70 MW/(m²·K·THz) in the classical approach and 64 MW/(m²·K·THz) in the quantum-corrected method. This indicates that vibrational modes in this frequency range play the most significant role in heat transfer across the interface. The phonon contribution to heat transport gradually approaches zero at approximately 20 THz, suggesting that higher-frequency phonons do not contribute significantly due to acoustic mismatch and strong scattering at the interface.

In contrast, the Si/4H-SiC interface exhibits a more complex spectral structure with multiple distinct peaks, as shown in Figure 3(c). The main peak occurs around 7 THz, with a maximum intensity of approximately 98 MW/(m²·K·THz), which is about 40% higher than the corresponding peak in 3C-SiC. A notable secondary peak appears near 14 THz with an intensity of about 89 MW/(m²·K·THz), and a weaker peak is also observed around 25 THz. This multi-peak structure reflects the anisotropic nature of the hexagonal 4H-SiC crystal structure, which leads to more distinct phonon branches and a broader range of active frequencies extending up to 40 THz. This spectral extension toward higher frequencies is one of the main reasons for the higher interfacial thermal conductance observed in 4H-SiC.

A deeper understanding of the relative contributions from different frequency ranges can be obtained through accumulated interfacial thermal conductance (Accumulated ITC) analysis. For the Si/3C-SiC interface, shown in Figure 3(b), the majority of the total thermal conductance is contributed by phonons with frequencies below 10 THz. The accumulated curve rises rapidly within the 0–10 THz range and then gradually approaches the final value of 613.38 MW/m²·K (quantum-corrected) or 746.24 MW/m²·K (classical). This behavior indicates that low-frequency acoustic phonons dominate heat transfer across the interface.

For the Si/4H-SiC interface, the accumulated curve shown in Figure 3(d) exhibits a different trend. Approximately 85% of the total thermal conductance is contributed by phonons with frequencies below 15 THz, indicating a broader participation of the frequency spectrum. The accumulated curve gradually increases up to 40 THz, reaching final values of 925.84 MW/m²·K (quantum-corrected) or 1244.06 MW/m²·K (classical). This value is about 51% higher than that of 3C-SiC, confirming the significant advantage of 4H-SiC for thermal management applications.

In the Si/3C-SiC interface, quantum corrections lead to an 18% reduction in ITC, decreasing the value from 746.24 to 613.38 MW/m²·K. In contrast, for Si/4H-SiC, the reduction is 26%, from 1244.06 to 925.84 MW/m²·K. This larger reduction in 4H-SiC arises from the substantial contribution of higher-frequency phonons, for which quantum effects such as the freezing of quantum degrees of freedom at temperatures below the Debye temperature—become more significant. At low frequencies (below 5 THz), the difference between the classical and quantum-corrected results is small, but as the frequency increases, the gap between the two curves widens considerably. These results carry important implications for the design of thermal systems. First, selecting the 4H-SiC polytype instead of 3C-SiC can substantially enhance interfacial thermal conductance. Comparison of the two structures shows that the Si/4H-SiC interface exhibits a higher ITC due to its broader range of active phonon modes and superior spectral overlap.

Second, the use of quantum-corrected calculations is essential for accurately predicting thermal performance, particularly at lower operating temperatures or in systems where high-frequency phonons play a significant role.

Third, understanding the spectral distribution of heat transport enables engineering of interfaces to optimize phonon matching within specific frequency ranges. Moreover, these calculations help provide a deeper understanding of the interfacial layer as a phonon bridge, clarifying how it mediates vibrational energy transfer across dissimilar materials.

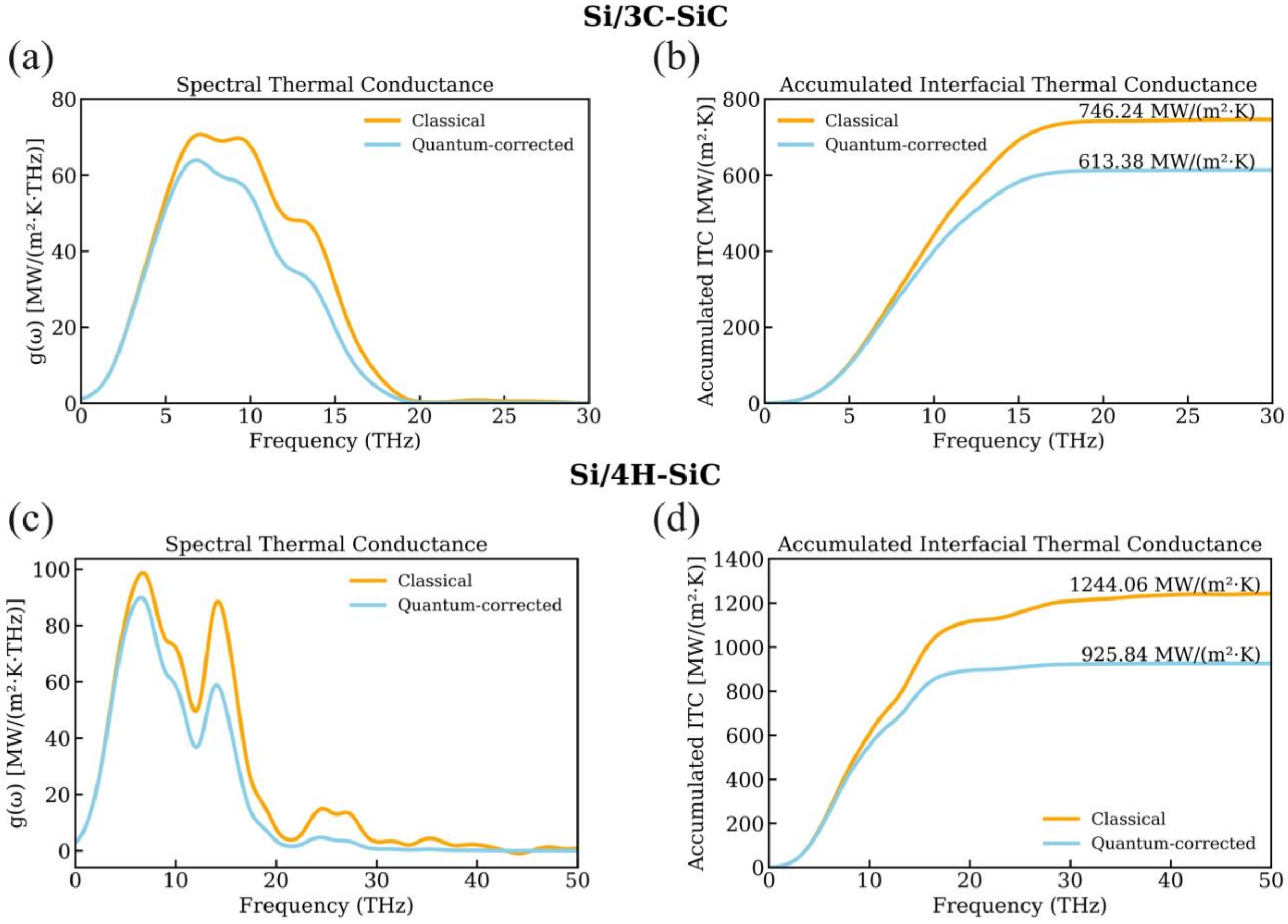

Figure 3: (a) Spectral interfacial thermal conductance of the Si/3C-SiC interface, (b) Accumulated interfacial thermal conductance of the Si/3C-SiC interface, (c) Spectral interfacial thermal conductance of the Si/4H-SiC interface, (d) Accumulated interfacial thermal conductance of the Si/4H-SiC interface.

### 3.2. Spectral and accumulated ITC analysis of SiC/diamond interfaces

Diamond is considered one of the most effective substrates for heat spreading in high-power electronics thermal management. As an ultra-wide bandgap (UWBG) semiconductor, it possesses the highest thermal conductivity known in nature. Its exceptionally high thermal conductivity and low dielectric constant make diamond an ideal candidate for use as a heat sink or within microchannel cooling systems in power devices, enabling a significant reduction in operating temperatures [53, 54].

The combination of diamond and SiC offers a significant improvement in heat dissipation efficiency due to its remarkably high thermal conductivity. To investigate this, we evaluated the ITC between diamond and two common SiC polytypes: 4H-SiC and 3C-SiC. In addition, the influence of amorphous interlayers composed of aSi, aSiC, and mixed amorphous C/SiC was systematically examined to assess their effectiveness in enhancing interfacial heat transport.

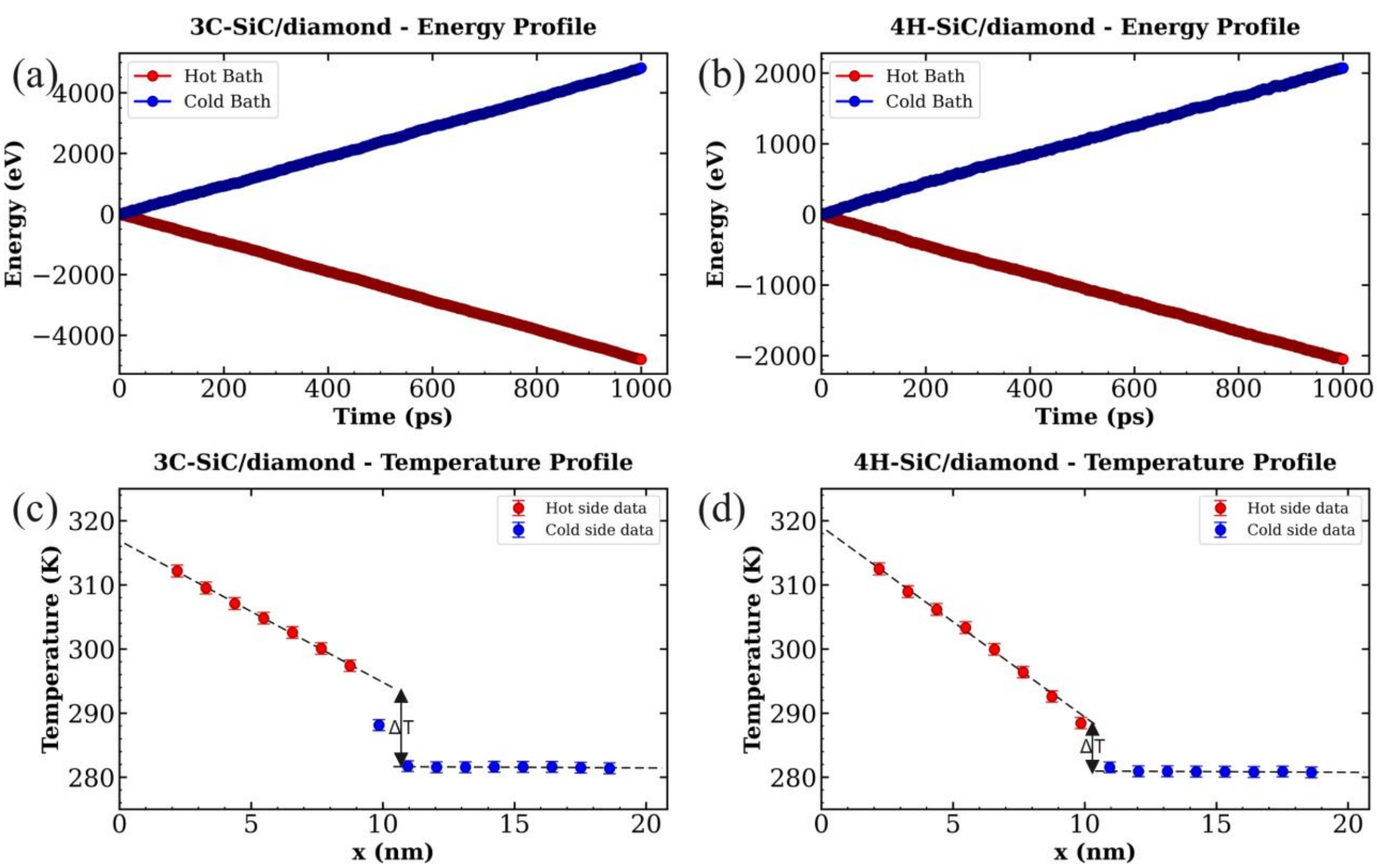

Figure 4. (a) Cumulative energy profile of the thermostats during NEMD simulations for the 3C-SiC/diamond heterostructure; (b) same for 4H-SiC/diamond; (c) steady-state temperature profile for the 3C-SiC/diamond heterostructure; (d) steady-state temperature profile for the 4H-SiC/diamond heterostructure.

The temperature profiles shown in Figures 4(c,d) indicate the presence of a distinct temperature drop near the SiC/diamond interface, reflecting the existence of an interfacial thermal resistance in this region. The most prominent feature of these profiles is the abrupt temperature jump at the interface. Additionally, the cumulative energy profiles show linear variations of the energy in the heat source and heat sink regions, confirming that the system has reached a steady state and that a constant heat flux has been established throughout the simulation.

Figures 5(a–d) illustrate the spectral interfacial thermal conductance, $g(\omega)$, and the accumulated interfacial thermal conductance for the 4H-SiC/diamond and 3C-SiC/diamond interfaces. These results include values obtained from classical molecular dynamics simulations as well as those corrected for quantum effects.

In the left panels, Figures 5(a) and 5(c), the spectral thermal conductance $g(\omega)$ is plotted as a function of phonon frequency. As observed, in both 4H-SiC/diamond and 3C-SiC/diamond interfaces, the main contribution to interfacial heat transfer is provided by low and mid-frequency phonons. Specifically, the largest contribution occurs in the approximate range of 5–30 THz, whereas phonons with frequencies higher than about 40 THz contribute only negligibly. This behavior indicates that heat transfer across these interfaces is primarily governed by acoustic modes and intermediate-frequency optical modes, while high-frequency phonons play a limited role due to spectral mismatch and strong scattering at the interface.

A comparison between the classical and quantum-corrected results shows that the value of $g(\omega)$decreases significantly across the entire frequency range after applying the quantum correction. This reduction is particularly pronounced in the high-frequency region. The reason is that classical simulations tend to overestimate the population of high-frequency phonons, whereas in the quantum description the excitation probability of these modes is lower at typical temperatures. Consequently, applying the quantum correction suppresses the contribution of high-frequency modes to heat transport and yields a more realistic distribution of phonon participation.

The right panels, Figures 5(b) and 5(d), show the accumulated interfacial thermal conductance as a function of frequency. These plots indicate that with increasing frequency, the contributions of phonon modes accumulate progressively until reaching the total interfacial thermal conductance. For the 4H-SiC/diamond interface, the accumulated ITC in the classical case is approximately 2169.47 $MW/m^2 \cdot K$, whereas the quantum-corrected value is about 1006.00 $MW/m^2 \cdot K$. In contrast, for the 3C-SiC/diamond interface, the classical accumulated ITC is approximately 2301.65 $MW/m^2 \cdot K$, and the quantum-corrected value is around 1121.15 $MW/m^2 \cdot K$, which is consistent with previously reported studies [28].

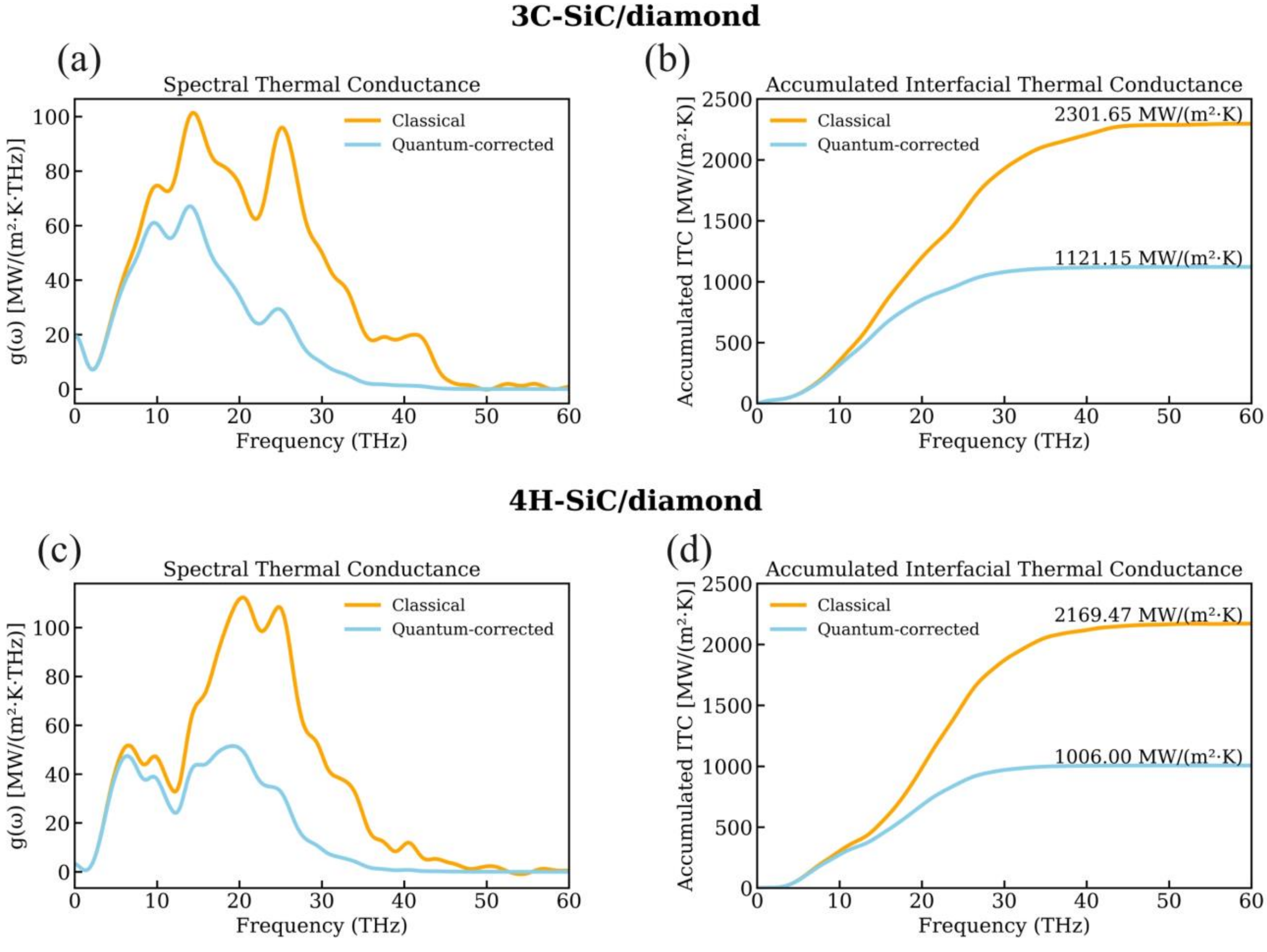

Figure 5. (a) Spectral interfacial thermal conductance of the 3C-SiC/diamond interface, (b) Accumulated interfacial thermal conductance of the 3C-SiC/diamond interface, (c) Spectral interfacial thermal conductance of the 4H-SiC/diamond interface, (d) Accumulated interfacial thermal conductance of the 4H-SiC/diamond interface.

A comparison of these results shows that the 3C-SiC/diamond interface exhibits a higher interfacial thermal conductance than the 4H-SiC/diamond interface. This difference can be attributed to variations in the crystal structure and the lattice dynamical properties of the two SiC polytypes. Owing to its simpler cubic symmetry, the crystal structure of 3C-SiC enables a more favorable phonon spectrum matching with diamond. This improved spectral compatibility enhances phonon spectral overlap, thereby facilitating more efficient interfacial heat transfer. The positive impact of good lattice matching on increasing ITC has also been confirmed in previous studies [28].

The ITC values were calculated for the sharp Si/SiC interface as well as for the Si/4H-SiC and Si/3C-SiC interfaces containing amorphous interlayers of aSi, aSi-aSiC, and aSiC .To investigate the role of the amorphous interlayer in heat transport across heterogeneous SiC/Si interfaces, we examined the effect of the thickness of these amorphous layers on the ITC of the interfaces at room temperature. The corresponding classical and quantum-corrected ITC values are presented in Figures 6(a–b).

As a reference, our results show that the ITC value for the sharp Si/3C-SiC interface is 613 $MW/m^2 \cdot K$ with quantum correction and 746 $MW/m^2 \cdot K$ without quantum correction. Similarly, for the sharp Si/4H-SiC interface, the ITC values with and without quantum correction are 926 $MW/m^2 \cdot K$ and 1244 $MW/m^2 \cdot K$, respectively.

### 3.3. Effect of amorphous interlayer type and thickness on ITC

A comparison between the classical and quantum-corrected calculations indicates that quantum effects play a significant role in interfacial heat transport. In Si-based structures, the classical ITC values are considerably higher than the quantum-corrected ones, with differences ranging from 18% to 26%. This discrepancy is more pronounced in 4H-SiC systems, suggesting that quantum effects become increasingly important in structures involving higher phonon frequencies. The quantum-corrected ITC values obtained in this study are consistent with previously reported simulation results [41, 55].

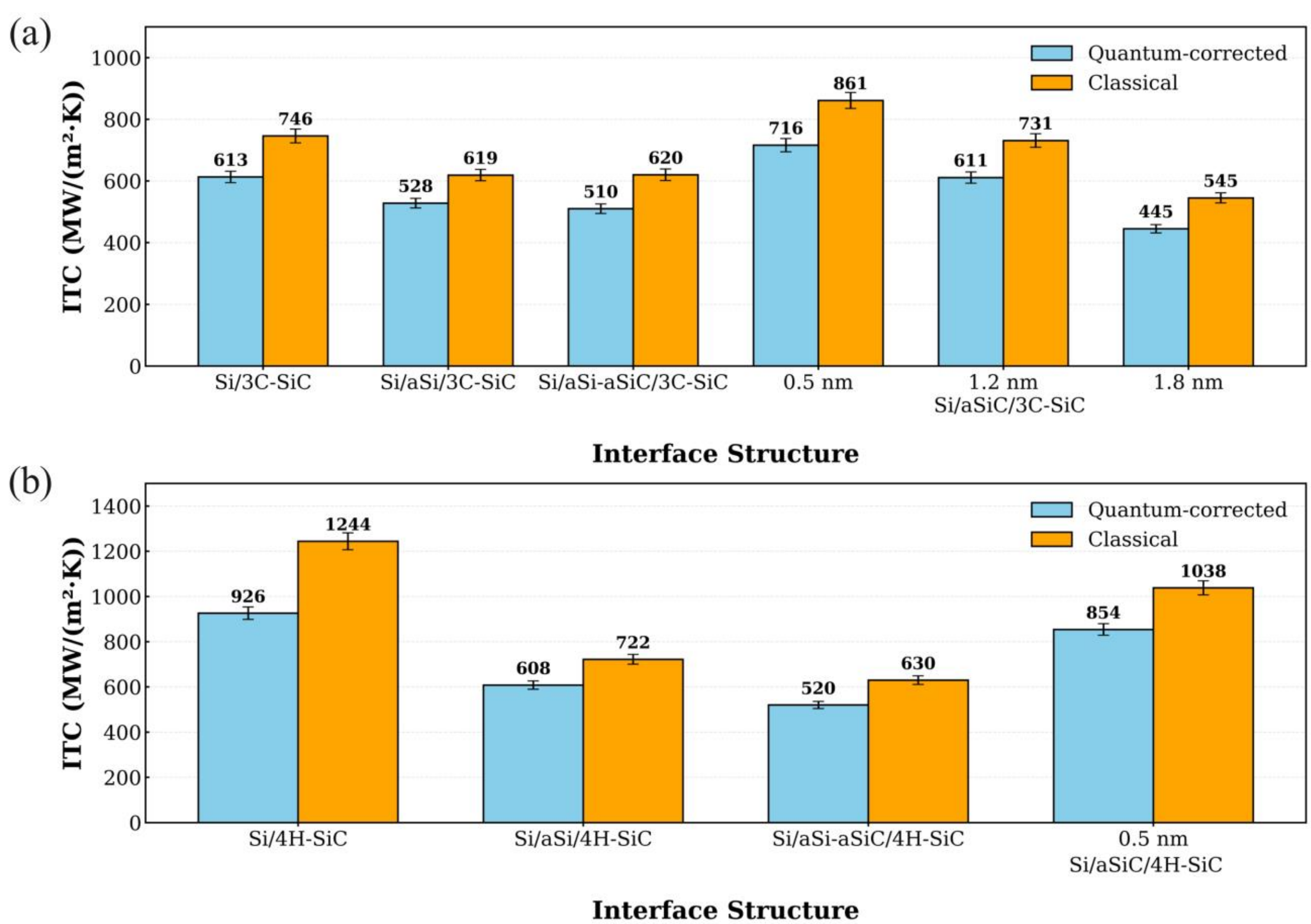

Figure 6: (a) Comparison of classical and quantum-corrected ITC values for the sharp Si/3C-SiC interface and interfaces containing amorphous interlayers with different thicknesses, including aSi, aSi-aSiC, and aSiC. (b) Comparison of classical and quantum-corrected ITC values for the sharp Si/4H-SiC interface and interfaces containing amorphous interlayers with different thicknesses, including aSi, aSi-aSiC, and aSiC.

Molecular dynamics simulation results indicate that the structure and thickness of amorphous interlayers have a significant influence on the ITC of Si/SiC systems. A comparison between the two SiC polytypes shows that the Si/4H-SiC interface exhibits considerably better thermal performance than Si/3C-SiC. The quantum-corrected ITC for the sharp Si/4H-SiC interface reaches 926 MW/m²·K, whereas for Si/3C-SiC it is only 613 MW/m²·K, corresponding to a 51% improvement in heat transfer.

Introducing amorphous layers at the Si/SiC interface produces complex effects on ITC that strongly depend on the type, thickness, and composition of the interlayer. In the Si/3C-SiC system, the introduction of a thin amorphous silicon layer reduces the ITC from 613 to 528 $MW/m^2 \cdot K$, corresponding to a 14% decrease. This reduction is attributed to increased inelastic phonon scattering within the disordered aSi structure. Similarly, in the Si/4H-SiC system, the presence of an aSi layer leads to a more significant reduction in ITC, from 926 to 608 $MW/m^2 \cdot K$, representing a 34% decrease. These results indicate that aSi layers act as thermal barriers and are therefore undesirable for thermal management applications.

The introduction of dual amorphous layers (aSi-aSiC) further degrades the thermal performance. In the Si/aSi–aSiC/3C-SiC structure, the ITC reaches 510 $MW/m^2 \cdot K$, whereas for Si/aSi–aSiC/4H-SiC it decreases to 520 $MW/m^2 \cdot K$. These correspond to changes of 17% and 44%, respectively, relative to the sharp interfaces, indicating that the addition of multiple amorphous layers cumulatively increases the overall thermal resistance of the system.

However, the use of controlled aSiC (0.5 nm) layers exhibits a different behavior. For a thin aSiC (0.5 nm) layer, the ITC in the Si/3C-SiC system increases to 716 $MW/m^2 \cdot K$, representing a 17% improvement compared with the sharp interface. This relative improvement can be attributed to the formation of intermediate vibrational states that may act as a "phonon bridge" between Si and 3C-SiC. In contrast, in the Si/4H-SiC system, a single aSiC layer results in an ITC of 854 $MW/m^2 \cdot K$, which is still 7.8% lower than the sharp interface value.

The effect of amorphous layer thickness on ITC is nonlinear and strongly dependent on the interlayer configuration. In the Si/3C-SiC system, the quantum-corrected ITC decreases from 716 $MW/m^2 \cdot K$ at 0.5 nm to 611 $MW/m^2 \cdot K$ at 1.2 nm, and further to 445 $MW/m^2 \cdot K$ at 1.8 nm — an overall reduction of approximately 38%. This behavior suggests that while a thin amorphous SiC interlayer may partially preserve phonon transmission, increasing thickness progressively enhances the intrinsic thermal resistance of the amorphous layer, which eventually dominates over any spectral bridging benefit. For the Si/4H-SiC system, thickness-dependent data were not pursued beyond 0.5 nm, as even this minimal interlayer thickness already leads to a substantial reduction in ITC relative to the crystalline Si/4H-SiC interface, confirming that the amorphous SiC layer acts as a thermal bottleneck regardless of the SiC polytype.

These findings have important implications for the design and optimization of SiC-based power electronic devices. Selecting the appropriate SiC polytype (4H-SiC versus 3C-SiC) can significantly influence thermal performance. Moreover, precise control of the interface structure and minimization of unintended amorphous layer formation during bonding or growth processes are essential for maintaining high ITC. If amorphous layers are unavoidable, employing ultra-thin a-SiC layers (less than ~1 nm) may provide a practical compromise between interface control and thermal performance.

The ITC values for SiC/diamond interfaces with 4H-SiC and 3C-SiC polytypes, together with aSi, mixed amorphous (aC-aSiC), and aSiC interlayers of 0.5 nm thickness, were also calculated. Both classical and quantum-corrected ITC values were extracted. As a reference, the ITC of the sharp 3C-SiC/diamond interface is 1121 $MW/m^2 \cdot K$ with quantum correction and 2302 $MW/m^2 \cdot K$ without quantum correction. Similarly, for the sharp 4H-SiC/diamond interface, the corresponding values are 1006 $MW/m^2 \cdot K$ and 2169 $MW/m^2 \cdot K$, respectively, as shown in Figure 7(a,b). The introduction of amorphous interlayers at the SiC/diamond interface has a much more detrimental effect compared with the Si/SiC system.

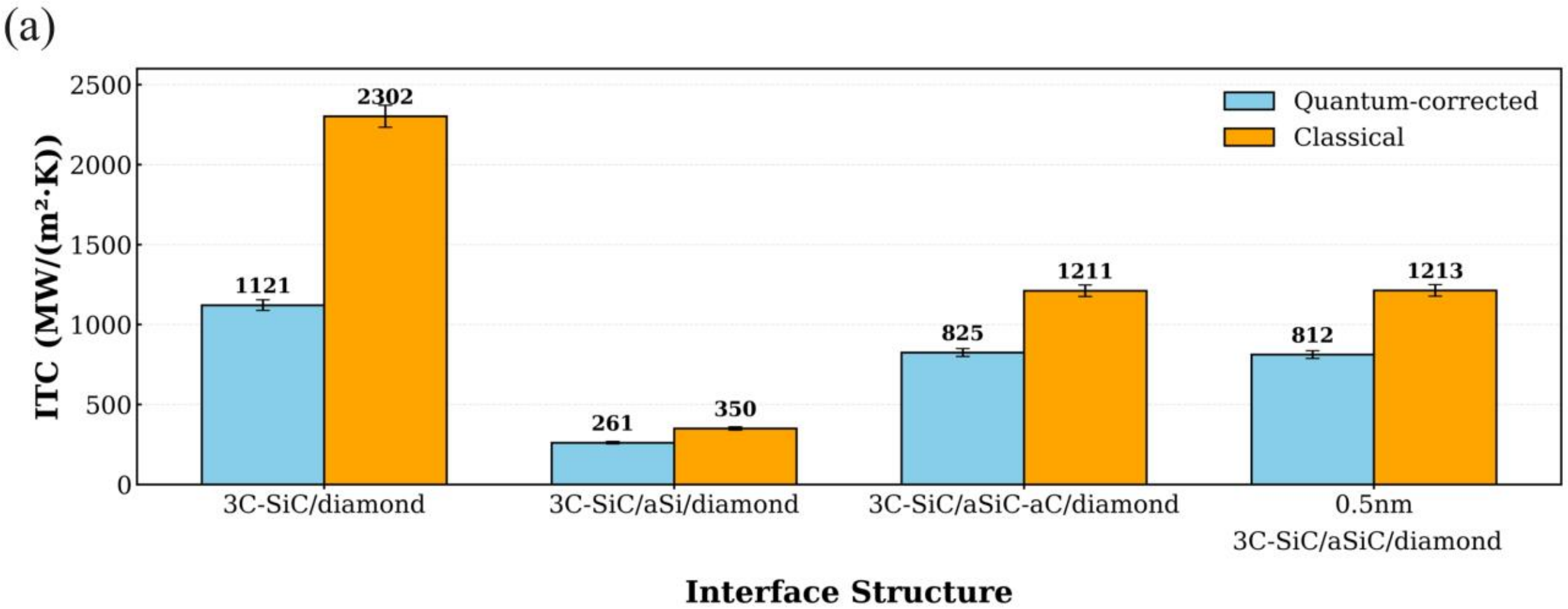

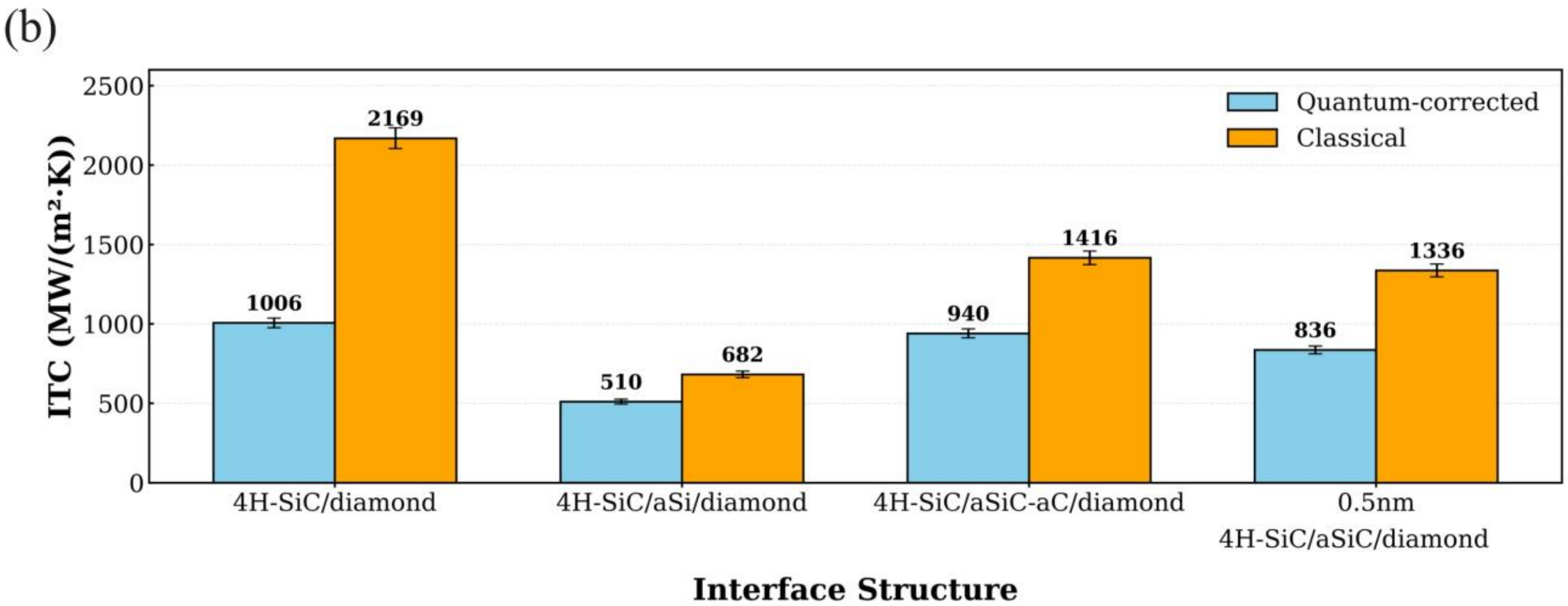

Figure 7: (a) Comparison of the classical and quantum-corrected ITC values for the sharp 3C-SiC/diamond interface and interfaces containing amorphous interlayers with different thicknesses, including aSi, mixed aC-aSiC, and aSiC layers. (b) Comparison of the classical and quantum-corrected ITC values for the sharp 4H-SiC/diamond interface and interfaces containing amorphous interlayers with different thicknesses, including aSi, mixed aC-aSiC, and aSiC layers

The introduction of a thin aSi interlayer causes a drastic reduction in thermal conductance. In the 3C-SiC/diamond system, the aSi layer reduces the ITC from 1121 to 261 MW/m²·K, corresponding to a 77% decline. In the 4H-SiC/diamond system, this layer decreases ITC by 49% (from 1006 to 510 MW/m²·K). For reference, the presence of aSi in Si/SiC systems caused only a 14–34% drop, indicating that aSi acts as a much stronger thermal barrier when interfacing with diamond.

The mixed amorphous structure (aC-aSiC) performs somewhat better than pure aSi but still reduces thermal conductance. In the 3C-SiC/diamond system, ITC decreases to 825 $MW/m^2 \cdot K$ (a 26% reduction relative to the sharp interface), while in 4H-SiC/diamond, ITC decreases slightly to 940 $MW/m^2 \cdot K$ (only 6.5% reduction).

In contrast to the Si/3C-SiC system, where adding an aSiC layer produced a “phonon-bridge” effect and improved thermal transfer by 17%, in SiC/diamond interfaces, even a very thin aSiC layer still causes a decrease in ITC. Specifically, in 3C-SiC/diamond, ITC drops by 27.6% to 812 $MW/m^2 \cdot K$, and in 4H-SiC/diamond, by 16.9% to 836 $MW/m^2 \cdot K$.

These results have significant implications for the design of thermal and electronic devices based on SiC/diamond interfaces, where diamond functions as a high-performance heat sink. The findings demonstrate that direct bonding (sharp interface) between 3C-SiC and diamond provides the highest heat transfer efficiency. Furthermore, during device fabrication and bonding processes, formation of amorphous interlayers — especially silicon-rich layers (aSi) — should be strictly avoided in order to fully exploit the exceptional thermal conductivity of diamond.

### 3.4. VDOS analysis, phonon bridge mechanism, and PPR

To gain deeper insight into the phonon transport mechanisms and the underlying origin of the observed variations in interfacial thermal conductance (ITC), the vibrational density of states (VDOS) of atoms located near the interface was analyzed. Figures 8(a) and 8(b) present the VDOS spectra for the sharp Si/3C-SiC interface and the Si/3C-SiC interface containing an amorphous interlayer, respectively.

As shown in Figure 8(a), the VDOS spectrum of Si atoms is mainly concentrated in the low-frequency range (0–20 THz), with its principal peak centered at around 15 THz. In contrast, the SiC spectrum extends to much higher frequencies (up to ~40 THz), with its main peaks located in the 25–35 THz range. This difference indicates a clear vibrational mismatch between Si and SiC. The shaded region in Figure 8 represents the spectral overlap, which is primarily confined to frequencies below 20 THz. To quantitatively evaluate vibrational compatibility between the two materials, the spectral overlap factor (S) between the VDOS of Si and SiC was calculated. Additional VDOS results for the other interfaces are provided in Section S3 of the Supporting Information. This parameter quantifies the overlap of vibrational states across the frequency spectrum and is defined as [41]:

$$S = \frac{\int VDOS_{SiC}(\omega) VDOS_{Si}(\omega)\, d\omega}{\int VDOS_{SiC}(\omega) d\omega \int VDOS_{Si}(\omega)\, d\omega} \quad (6)$$

where $VDOS_{SiC}(\omega)$and $VDOS_{Si}(\omega)$represent the vibrational density of states of SiC and Si, respectively. A larger value of S indicates a greater spectral overlap between the phonon states of the two materials and therefore a better vibrational matching at the interface.

As shown in Fig. 8(a), the sharp interface exhibits a small overlap factor of $S = 0.012$, indicating limited vibrational overlap between the Si and SiC spectra. After introducing the aSiC interlayer, this value increases to $S = 0.026$, indicating improved spectral matching and enabling additional phonon transport channels. This behavior originates from the broadening of the VDOS spectrum of the amorphous layer over a frequency range extending from near zero to approximately 60 THz. As seen in Fig. 8(b), the green curve corresponding to the amorphous layer shows substantial overlap with Si at low frequencies and with SiC at high frequencies, thereby acting as a “phonon bridge.” Since Si exhibits negligible vibrational density of states above 20 THz, high-frequency phonons in SiC are predominantly scattered or reflected at the direct interface. However, the presence of the amorphous interlayer introduces intermediate vibrational modes that reduce the intrinsic vibrational mismatch and thereby enhance interfacial heat transport.

The phonon-bridge enhancement observed in the Si/3C-SiC interface can be rationalized based on the VDOS spectral overlap analysis. As previously mentioned, the spectral overlap factor increases from S = 0.012 for the sharp interface to S = 0.026 after introducing the ultrathin aSiC interlayer, indicating a more than twofold increase in vibrational coupling. The broadened vibrational spectrum of amorphous SiC introduces continuous intermediate-frequency modes within the range (~5–15 THz), which overlap with both materials' phonons and enable additional transmission pathways. Because the aSiC layer is extremely thin (0.5 nm), the additional thermal resistance associated with structural disorder remains limited, allowing the improvement in spectral coupling to outweigh disorder-induced phonon scattering. However, increasing the thickness of the amorphous layer introduces greater intrinsic thermal resistance, which can ultimately lead to a reduction in ITC [40]. In contrast, the sharp Si/4H-SiC interface already exhibits broader intrinsic phonon participation and stronger spectral coupling at higher frequencies; therefore, the additional vibrational states introduced by the amorphous layer provide only limited benefit, while disorder-induced scattering becomes dominant. For SiC/diamond interfaces, where heat transport strongly depends on high-frequency phonons, the amorphous layer induces strong scattering of short-wavelength phonons and therefore contributes mainly to interfacial thermal resistance.

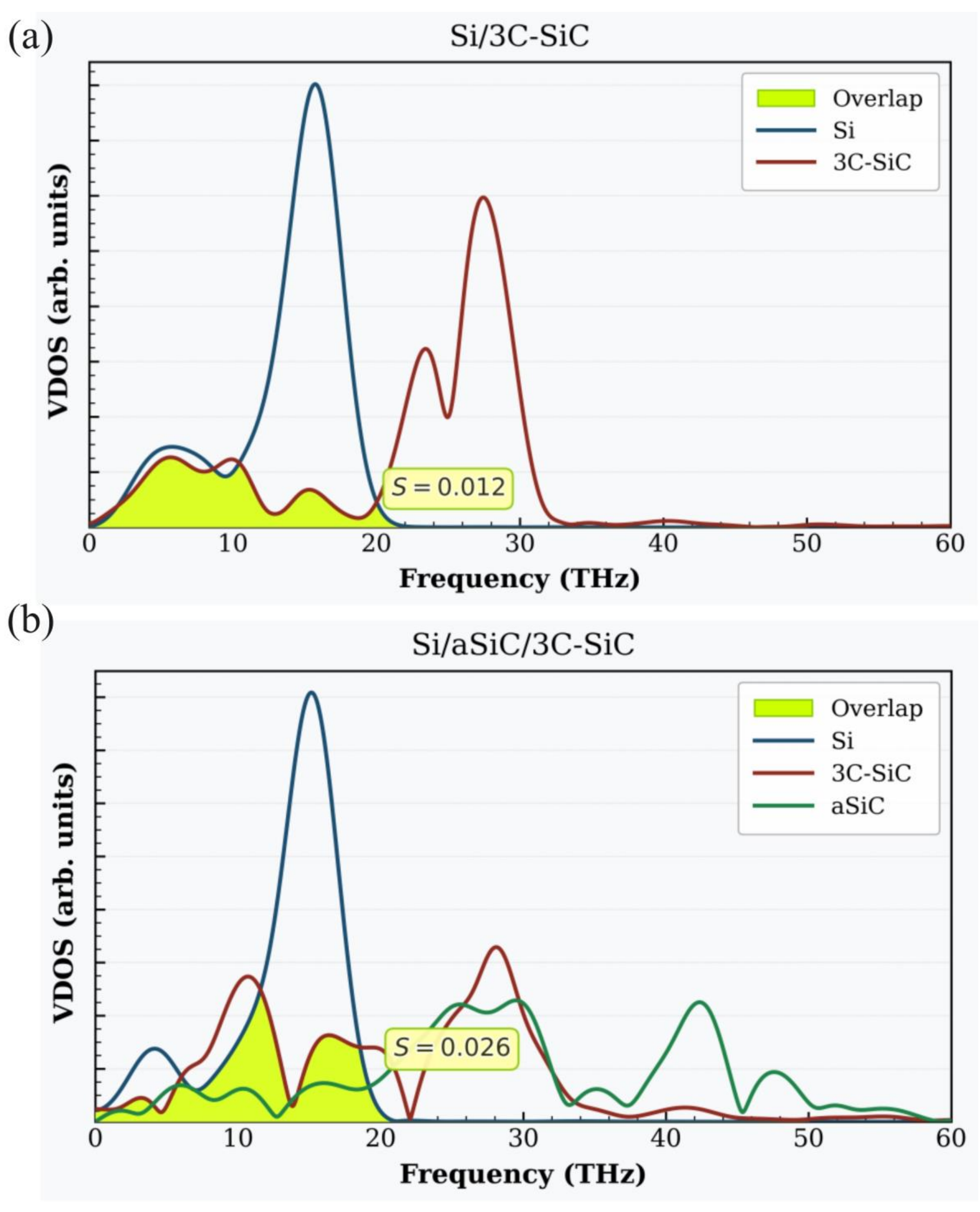

Figure 8: (a) VDOS for the sharp Si/3C-SiC interface, (b) VDOS for the Si/aSiC (0.5 nm)/3C-SiC interface.

The different roles of the ultrathin aSiC interlayer in Si/3C-SiC, Si/4H-SiC, and SiC/diamond interfaces originate from the competition between two effects: phonon spectral bridging and disorder-induced phonon scattering. In the Si/3C-SiC system, the sharp interface exhibits a relatively limited spectral overlap because the Si vibrational spectrum is mainly confined below ~20 THz, whereas 3C-SiC supports higher-frequency vibrational modes [55]. Introducing an ultrathin 0.5 nm aSiC layer broadens the interfacial vibrational spectrum and creates intermediate phonon states that partially bridge the spectral mismatch, leading to a slight ITC enhancement. However, in the Si/4H-SiC system, the sharp interface already possesses stronger intrinsic phonon coupling and broader phonon participation due to the complex hexagonal phonon branches of 4H-SiC. Therefore, the additional spectral-bridging benefit introduced by the amorphous layer becomes insufficient to compensate for the enhanced phonon scattering caused by structural disorder. In SiC/diamond interfaces, the disorder effect becomes dominant because heat transfer strongly depends on efficient transmission of high-frequency phonons associated with diamond. Consequently, the amorphous aSiC layer suppresses high-frequency phonon transport and acts primarily as a thermal resistance layer rather than a phonon bridge.

To further characterize phonon transport behavior at the interface, the PPR was employed to evaluate the degree of phonon propagation and localization over different frequency ranges [28, 56]:

$$\mathrm{PPR}(\omega) = \frac{1}{\mathrm{N}} \frac{\left(\sum_i DOS(\omega)_i^2\right)^2}{\left(\sum_i DOS(\omega)_i^4\right)} \tag{7}$$

where N is the total number of atoms, and $DOS\ (\omega)_i$ represents the phonon density of states of the $i$-th atom obtained from the Fourier transform of its velocity autocorrelation function. The PPR quantitatively describes the spatial extent of vibrational modes within the system. Values on the order of $O(1)$ correspond to extended (delocalized) propagating modes that can effectively contribute to thermal transport, whereas values approaching $O(\frac{1}{N})$ indicate strongly localized vibrational modes with limited participation in heat conduction [57].

Figure 9(a, b) compares the PPR for the sharp Si/3C-SiC interface and the Si/aSiC (0.5 nm)/3C-SiC system. PPR values close to unity correspond to highly delocalized heat-carrying phonons, whereas values approaching zero indicate localized vibrational modes that suppress thermal transport.

For the sharp interface, the Si PPR exhibits pronounced oscillations within the intermediate-frequency range, decreasing to approximately 0.45 near 11 THz. These fluctuations indicate phonon confinement effects near the interface, which partially localize vibrational modes that would otherwise remain extended in bulk Si. After introducing the 0.5 nm amorphous SiC interlayer, these oscillations are significantly suppressed, and the Si PPR remains relatively stable between 0.84 and 0.96 throughout the ~4–18 THz frequency range. This behavior indicates that the ultrathin amorphous layer helps preserve the delocalized nature of intermediate-frequency phonons near the interface. Simultaneously, the PPR of 3C-SiC within the same frequency range increases from approximately 0.25–0.40 to about 0.32–0.63, suggesting improved spatial delocalization and stronger participation of SiC vibrational modes in interfacial heat transport.

Importantly, both materials exhibit higher PPR values within the same intermediate-frequency region after the insertion of the amorphous layer. This indicates that a larger number of spatially extended and transport-active phonon modes become active simultaneously for vibrational coupling across the interface. Such behavior supports the phonon-bridge mechanism and explains the enhancement of ITC from 613 to 716 $MW/m^2 \cdot K$.

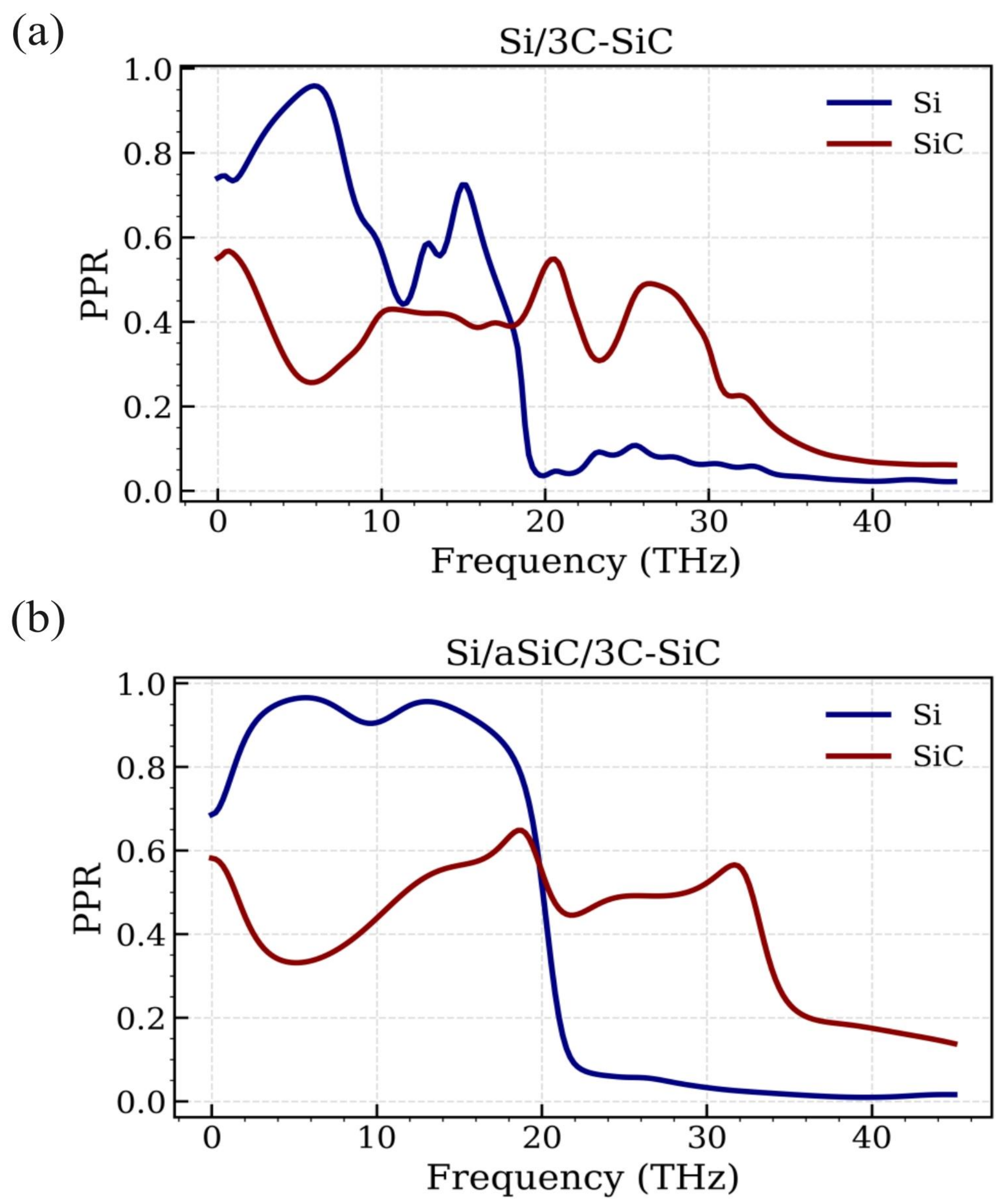

Figure 9: Comparison of the PPR for (a) the sharp Si/3C-SiC interface and (b) the Si/aSiC (0.5 nm)/3C-SiC system.

### 3.5. Temperature dependence of ITC

The operating temperature of electronic devices is a key factor that strongly affects their reliability and thermal performance. Therefore, investigating the temperature dependence of heat transfer across various semiconductor interfaces has substantial practical importance. To understand the thermal behavior of these junctions, the temperature dependence of the ITC was examined over the range 200–500 K for five different configurations. Figure 10 presents the ITC variation for silicon-based structures Si/3C-SiC, Si/4H-SiC, and the amorphous-interlayered interface (Si/aSiC(0.5 nm)/3C-SiC) as well as for diamond-based heterostructures: 3C-SiC/diamond and 4H-SiC/diamond.

As seen in Figure 10, all five configurations exhibit an increase in ITC with temperature from 200 K to 500 K. The Si/4H-SiC interface (purple squares) exhibits the highest ITC among Si-based interfaces, rising from approximately 653 $MW/m^2 \cdot K$ at 200 K to about 1328 $MW/m^2 \cdot K$ at 500 K. In contrast, the direct Si/3C-SiC interface (blue circles) shows the lowest ITC—about 490 $MW/m^2 \cdot K$ at 200 K increasing to 658 $MW/m^2 \cdot K$ at 500 K—nearly saturating at higher temperatures. However, inserting an ultrathin 0.5 nm amorphous SiC layer between Si and 3C-SiC (orange triangles) significantly enhances heat transport, elevating ITC to around 1083 $MW/m^2 \cdot K$ at 500 K. This confirms the "phonon bridge" role of the amorphous layer.

On the other hand, the diamond-based heterostructures demonstrate a significant increase. The 4H-SiC/diamond interface (red downward triangles) shows the highest ITC among all diamond-based interfaces, rising steeply and nonlinearly from roughly 721 $MW/m^2 \cdot K$ at 200 K to exceed 2751 $MW/m^2 \cdot K$ at 500 K. The 3C-SiC/diamond interface (green diamonds), although initially comparable to the Si/SiC interfaces at 200 K (~514 $MW/m^2 \cdot K$), reaches a high 2029 $MW/m^2 \cdot K$ at 500 K due to stronger spectral coupling with diamond and activation of high-energy phonons.

The increase in ITC with temperature is mainly attributed to enhanced phonon excitation and enhanced phonon scattering at elevated temperatures. As temperature rises, more mid- and high-frequency phonons become thermally activated and participate in interfacial heat transport, providing additional channels for energy transfer across the interface, a behavior also reported in previous MD studies of solid–solid interfaces [40, 58, 59]. This effect is particularly significant in SiC/diamond interfaces because diamond exhibits abundant high-frequency phonon modes and an extremely high Debye temperature, so higher temperatures activate more energetic phonons and lead to a more pronounced increase in ITC compared with Si/SiC systems. For interfaces containing amorphous interlayers, however, the temperature dependence of ITC results from the competition between thermally enhanced phonon transport and disorder-induced scattering. While moderate temperature increases can activate additional vibrational modes and strengthen anharmonic phonon interactions within the amorphous region, further temperature increases excite more short-wavelength phonons that are strongly scattered by the structural disorder of the amorphous layer, thereby increasing the interfacial thermal resistance and weakening the overall enhancement of ITC compared with sharp crystalline interfaces.

This temperature-dependent analysis indicates that, for high-power electronic devices operating at elevated temperatures, replacing silicon with diamond is essential for efficient thermal management. Moreover, near diamond interfaces, the hexagonal 4H-SiC structure exhibits a distinct advantage over 3C-SiC.

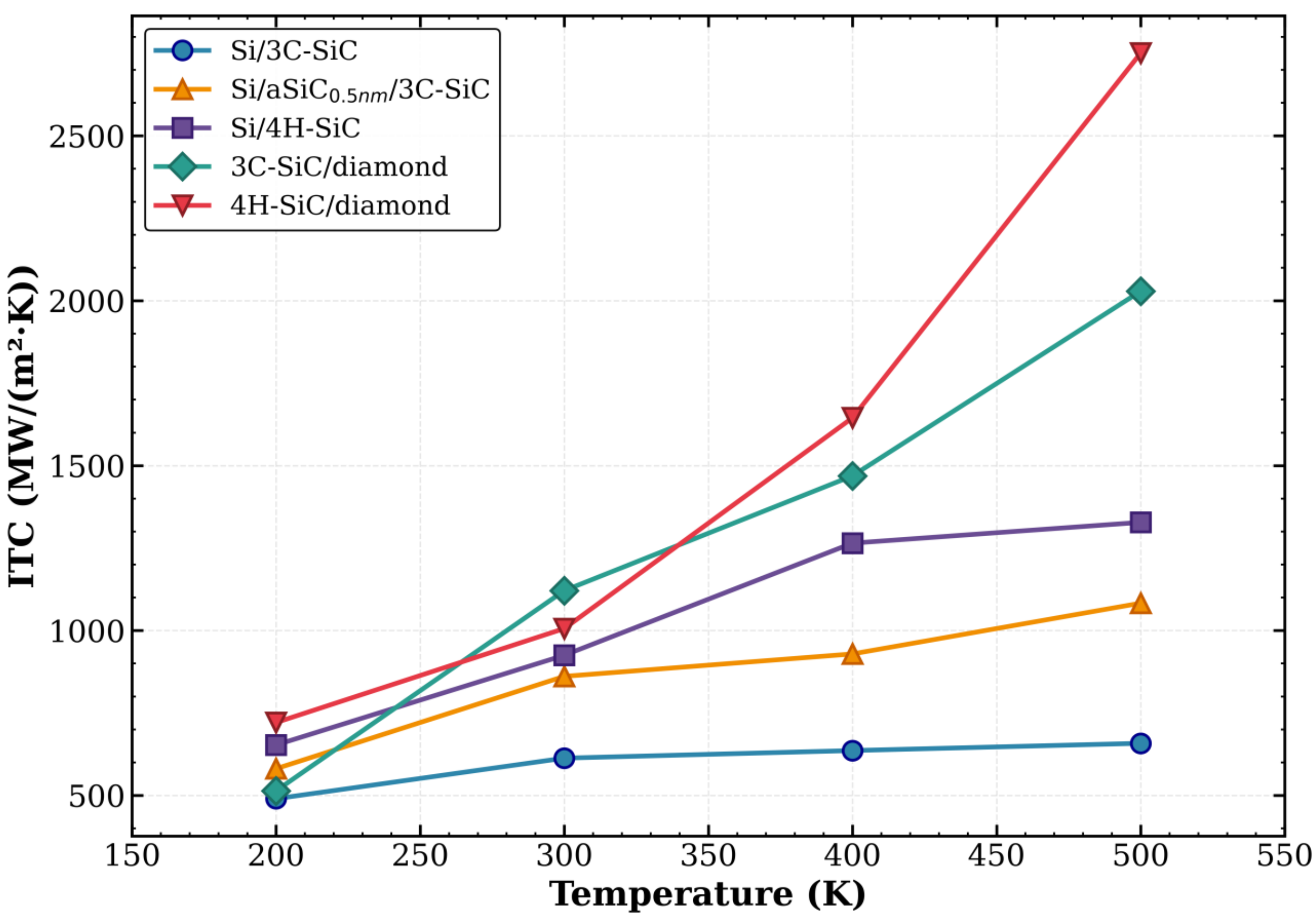

Figure 10: Temperature dependence of the ITC for silicon-based interfaces (Si/3C-SiC, Si/4H-SiC, and Si/aSiC (0.5 nm)/3C-SiC) and diamond-based heterostructures (3C-SiC/diamond and 4H-SiC/diamond) over the temperature range of 200–500 K.

### 3.6. Comparison with experimental and theoretical literature

In this section, the results are categorized into three groups: experimental data, theoretical predictions, and the results of the present study. Figure 11 (a–b), a comprehensive and visual comparison of ITC values for two groups of heterogeneous systems: (a) Si/SiC and, (b) SiC/diamond.

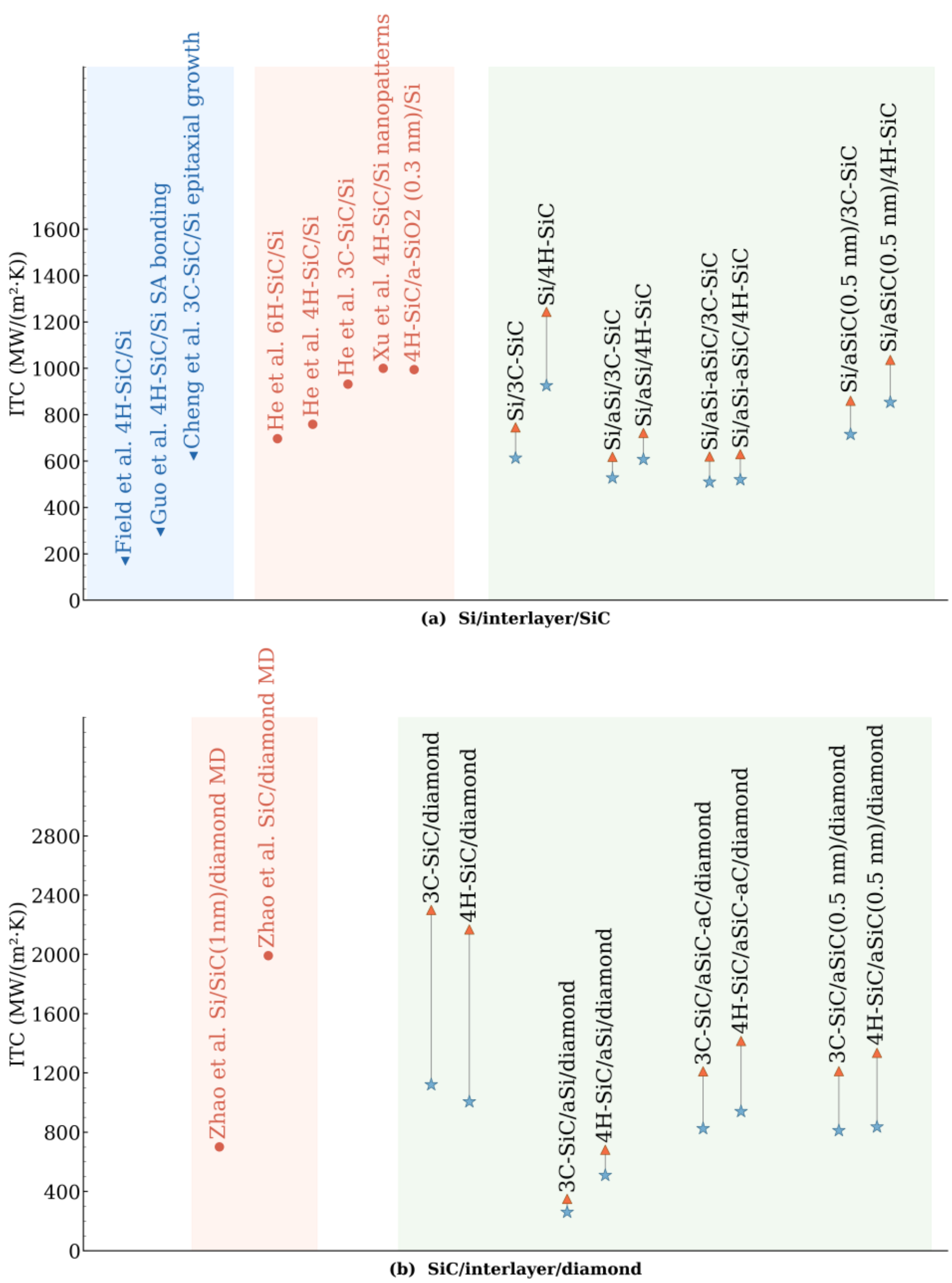

Figure 11: Comparison of the ITC values for various (a) Si/SiC and, (b) SiC/diamond interfaces, using experimental data [27, 34, 36] and theoretical calculations [28, 33, 41, 55].

Field et al. [36], investigating a bonded 4H-SiC/Si interface, reported the lowest ITC value of 167 MW/m²·K. Guo et al. [34], using the surface activated bonding (SAB) technique for 4H-SiC/Si, obtained a value of about 293 MW/m²·K. The highest experimental value was reported by Cheng et al. [27] for epitaxially grown 3C-SiC/Si, reaching 620 MW/m²·K. The quantum-corrected value obtained in the present study for the sharp 4H-SiC/Si interface (926 MW/m²·K) is significantly higher than the experimental values reported by Field and Guo, highlighting the detrimental impact of interfacial defects and imperfections in real systems. However, the experimental result of Cheng et al. for 3C-SiC (620 MW/m²·K) shows excellent agreement with the quantum-corrected result of the present study for the sharp 3C-SiC/Si interface (613 MW/m²·K).

Among theoretical studies, He et al. [55] reported ITC values for several SiC polytypes: 697 MW/m²·K for 6H-SiC/Si, 759 MW/m²·K for 4H-SiC/Si, and 932 MW/m²·K for 3C-SiC/Si. Xu et al. [33] obtained a relatively high value of about 1000 MW/m²·K for nanopatterned 4H-SiC/Si structures. Another theoretical work involving an interlayer structure 4H-SiC/a$SiO_2$ (0.3 nm)/Si reported an ITC of 994.7 MW/m²·K [41].

In this comparative analysis, the quantum-corrected values of the present study fall within a similar range to previous theoretical studies, confirming the reliability of the present modeling approach.

In Figure 11(b), the comparison is limited to theoretical studies for the SiC/diamond system. The significantly larger vertical axis scale highlights the superior thermal transport capability of diamond as a heat-sink material.

Zhao et al. [28] reported an ITC of about 701.1 MW/m²·K for a Si/crystal SiC (1 nm)/diamond structure with a thin SiC interlayer. In another simulation, Zhao et al. predicted a very high ITC of about 1992.6 MW/m²·K for a sharp SiC/diamond interface.

Furthermore, our analysis indicates that replacing a silicon heat sink with diamond leads to an approximately 83 % increase in ITC, rising from about 613 MW/m²·K for the sharp Si/3C-SiC interface to 1121 MW/m²·K for the 3C-SiC/diamond interface. This remarkable improvement demonstrates that forming a sharp, amorphous-free interface in SiC/diamond structures can significantly enhance heat-transfer capability compared with conventional silicon heat sinks and several previously reported theoretical models, representing a valuable advancement for thermal management in high-power electronic devices.

## 4. Conclusion

This study employed NEMD simulations combined with spectral heat current (SHC), PPR, and VDOS analyses to investigate phonon-mediated interfacial thermal transport in Si/SiC and SiC/diamond heterostructures. The effects of SiC polytype, amorphous interlayer type and thickness, and quantum corrections were systematically examined.

Polytype selection strongly influences the baseline ITC. For Si-based interfaces, the quantum-corrected ITC of Si/4H-SiC (926 $MW/m^2 \cdot K$) is 51% higher than that of Si/3C-SiC (613 $MW/m^2 \cdot K$), owing to the broader phonon spectrum and enhanced mid-frequency phonon participation in the hexagonal structure. In contrast, at SiC/diamond interfaces, 3C-SiC provides better phonon spectral matching with diamond, resulting in an ITC of 1121 $MW/m^2 \cdot K$, approximately 11% higher than that of 4H-SiC/diamond (1006 $MW/m^2 \cdot K$). These results demonstrate that the optimal SiC polytype depends strongly on the adjacent material and interface configuration.

Amorphous interlayers exhibit strongly system-dependent effects. In Si/3C-SiC, an ultrathin 0.5 nm aSiC interlayer increases ITC by approximately 17% compared with the sharp interface. VDOS analysis shows that the spectral overlap factor increases from $S=0.012$ to $S=0.026$, indicating a phonon bridge effect that mitigates the vibrational mismatch between Si and SiC. However, increasing the aSiC thickness to 1.8 nm results in a substantial monotonic reduction in ITC due to the intrinsic thermal resistance of the amorphous layer. In contrast, aSi consistently acts as a thermal barrier, reducing ITC by approximately 14–34% in Si/SiC systems and by 49–77% in SiC/diamond interfaces, highlighting the importance of atomically sharp bonding when utilizing diamond substrates.

Quantum corrections are essential for reliable predictions of interfacial thermal transport. Classical molecular dynamics overestimates ITC by approximately 18–26% in Si/SiC. In SiC/diamond interfaces, the discrepancy becomes significantly larger, with classical predictions reaching more than twice the quantum-corrected values. This behavior originates from the excessive population of high-frequency phonons in classical statistics, particularly in high-Debye-temperature materials such as diamond (>2000 K) and SiC (~1200–1300 K).

Replacing Si with diamond in the heterostructure significantly enhances thermal transport. The ITC increases by approximately 83%, from 613 $MW/m^2 \cdot K$ for Si/3C-SiC to 1121 $MW/m^2 \cdot K$ for 3C-SiC/diamond. Temperature-dependent simulations further show a monotonic increase in ITC with increasing temperature, indicating enhanced inelastic phonon transport at elevated temperatures.

Based on these findings, several interface design guidelines can be proposed: employ 4H-SiC for Si-based interfaces, use 3C-SiC for diamond substrates, introduce an ultrathin (~0.5 nm) aSiC interlayer at Si/3C-SiC interfaces to exploit the phonon bridge effect, and avoid amorphous silicon formation, particularly at SiC/diamond interfaces.

Overall, the proposed theoretical framework provides valuable physical insight into phonon transport engineering and may contribute to improved thermal management strategies for next-generation high-power electronic devices.

## Declaration of competing interest

The authors declare that they have no known competing financial interests or personal relationships that could have appeared to influence the work reported in this paper.

## Acknowledgment

This work is based upon research funded by Iran National Science Foundation (INSF) under project No.4042701.

### CRediT Author Statement

**Pedram Mirchi:**

Conceptualization, Methodology, Software, Formal analysis, Investigation, Writing – Original Draft, Visualization

**Shirin Sabokdast:**

Conceptualization, Writing – Original Draft, Writing – Review & Editing

**Ali Rajabpour:**

Conceptualization, Supervision, Project administration, Funding acquisition, Writing – Review & Editing